\documentclass[%
 reprint,
superscriptaddress,
 amsmath,amssymb,
 aps,
 prl,
]{revtex4-1}

\usepackage{graphicx}% Include figure files
\usepackage{dcolumn}% Align table columns on decimal point
\usepackage{bm}% bold math
\usepackage[utf8]{inputenc}
\usepackage[T1]{fontenc}
\usepackage{verbatim}
\usepackage{units}
\usepackage{mathtools}
\usepackage{wasysym}
\usepackage{siunitx}
\usepackage{xfrac}
\usepackage{xcolor}
\usepackage[colorlinks, citecolor={blue!50!black}, urlcolor={blue!50!black}, linkcolor={red!50!black}]{hyperref}
\usepackage{bookmark}
\usepackage{tabularx}
\usepackage{microtype}
\usepackage[american]{babel}
\usepackage{cleveref}
\usepackage{soul}
\usepackage{xcolor}

\hypersetup{pdfauthor={D. Szombati et al.}, pdftitle={Quantum rifling: protecting a qubit from measurement back-action}} 
\setstcolor{blue}
\begin{document}

%\preprint{APS/123-QED}

\title{Quantum rifling: protecting a qubit from measurement back-action}% Force line breaks with \\

\author{Daniel Szombati}
 \altaffiliation[Current address:]{daniel.szombati@ens-lyon.fr}
 
\affiliation{ARC Centre of Excellence for Engineered Quantum Systems, Queensland 4072, Australia}
\affiliation{School of Mathematics and Physics, University of Queensland, St Lucia, Queensland 4072, Australia}

\author{Alejandro Gomez Frieiro}
\affiliation{ARC Centre of Excellence for Engineered Quantum Systems, Queensland 4072, Australia}
\affiliation{School of Mathematics and Physics, University of Queensland, St Lucia, Queensland 4072, Australia}

\author{Clemens M\"uller}
\affiliation{IBM Research Z\"urich, 8803 R\"uschlikon, Switzerland}
%\affiliation{Institute for Theoretical Physics, ETH Zürich, Switzerland}

\author{Tyler Jones}
\affiliation{ARC Centre of Excellence for Engineered Quantum Systems, Queensland 4072, Australia}
\affiliation{School of Mathematics and Physics, University of Queensland, St Lucia, Queensland 4072, Australia}

\author{Markus Jerger}
\affiliation{ARC Centre of Excellence for Engineered Quantum Systems, Queensland 4072, Australia}
\affiliation{School of Mathematics and Physics, University of Queensland, St Lucia, Queensland 4072, Australia}

\author{Arkady Fedorov}
\affiliation{ARC Centre of Excellence for Engineered Quantum Systems, Queensland 4072, Australia}
\affiliation{School of Mathematics and Physics, University of Queensland, St Lucia, Queensland 4072, Australia}

\date{\today}% It is always \today, today,

\begin{abstract}
Quantum mechanics postulates that measuring the qubit's wave function results in its collapse,  with the recorded discrete outcome designating the particular eigenstate the qubit collapsed into. 
We show this picture breaks down when the qubit is strongly driven during measurement.
More specifically, for a fast evolving qubit the measurement returns the time-averaged expectation value of the measurement operator, erasing information about the initial state of the qubit, while completely suppressing the measurement back-action.
We call this regime “quantum rifling”, as the fast spinning of the Bloch vector protects it from  deflection into either of its two eigenstates. We study this phenomenon with two superconducting qubits coupled to the same probe field and demonstrate that quantum rifling allows us to measure either one of the two qubits on demand while protecting the state of the other from measurement back-action. 
Our results allow for the implementation of selective read out multiplexing of several qubits, contributing to efficient scaling up of quantum processors for future quantum technologies.
\end{abstract}

\maketitle

%\tableofcontents

\begin{figure*}[ht]
    \centering
    \includegraphics[scale=0.35]{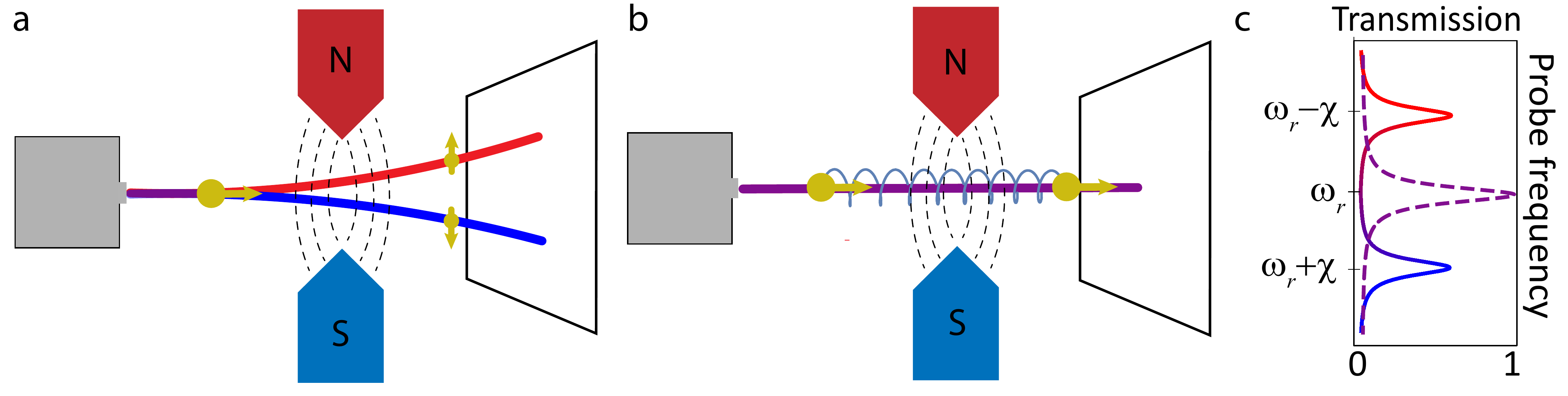}
    \caption{\textbf{Quantum rifling in a Stern-Gerlach setup. a,} Measurement of a spin initially prepared in a superposition state. The magnetic field measures the spin by deflecting its flight path in a direction dependent on the measured spin orientation. \textbf{b,} Measurement during rifling; when the spin is stronlgy driven during measurement the magnetic field cannot discriminate between spin states and the spin's flight path is undisturbed. \textbf{c,} Detection mechanism of the qubit state in our setup; the resonator experiences a qubit state-dependent shift in its resonance frequency by  $\pm\chi$ relative to its bare frequency $\omega_r$ in the rotating frame. Thus after averaging the results of experiment (a) one would observe the red and blue resonator transmission curves. During quantum rifling, when the qubit's state is driven resonantly with its transition energy, the resonator's resonance shifts to $\omega_r$, leading to the dashed transmission curve following the averaging of the experiment presented in (b).
      }
        \label{fig:sterngerlach}
\end{figure*}

The Stern-Gerlach experiment, originally conducted to demonstrate quantization in atomic-scale systems\cite{Gerlach1922}, is the prototypical example of a quantum measurement with a linear detector: an electron (or qubit) flying through a magnetic field is deflected from its straight path, with probabilities dependent on the qubit's initial state (see Figure  \ref{fig:sterngerlach}a). 
The measurement projects the state of the qubit onto either of its two spin eigenstates: $\pm \hbar/2$.

Driving the qubit with Rabi frequency $\Omega_R$ during the measurement leads to competition between the state evolution and the measurement projection. Such a scenario has been studied theoretically~\cite{Milburn1988,blais2004cavity,Gambetta2008} and experimentally~\cite{Hatridge2013, Ficheux2018} in the  strong measurement regime $\Omega_R\ll\Gamma_m$, where $\Gamma_m$ is the measurement rate at which information is extracted from the qubit. 
This regime is commonly described by the Quantum Zeno effect\cite{Itano1990}: a strong quantum measurement freezes the qubit's state, with occasional transitions occurring as sudden quantum jumps\cite{Schulman1998,Streed2006,Vijay2011,Slichter2016,minev2019catch} with rate $\propto\Omega_R^2/\Gamma_m$. 

The regime of strong driving $\Omega_R\gg\Gamma_m$, referred to as the sub-Zeno limit, has attracted attention in the context of continuous weak measurements\cite{Stace2004}. When the probe's bandwidth $\delta \omega$ exceeds the Rabi frequency, $\Omega_R<\delta\omega$, signatures of coherent Rabi oscillations appear in the detector signal~\cite{Goan2001, Korotkov2001, Korotkov2001a, Gurvitz2003}. 
It has been shown that the back-action introduced by the measurement imposes a fundamental limit on  the detection of oscillations and can be used to determine the quantum efficiency of the detector\cite{Korotkov2001,jordan2005continuous,Jordan2006}, or even to test the Leggett-Garg inequality\cite{Jordan2006a,Palacios-Laloy2010}. The opposite limit  $\Omega_R>\delta\omega$, where the Rabi frequency exceeds the bandwidth, is suitably described by the average Hamiltonian theory~\cite{Maricq1982}.  This regime however, has not yet been investigated in the context of continuous qubit measurement neither theoretically nor experimentally.

In this Letter, we study the measurement of a continuously driven qubit in the regime where the Rabi frequency dominates all other relevant parameters: $\Omega_R\gg\Gamma_m, \delta\omega$. First, we show that when the probe's bandwidth is not sufficient to follow the qubit's state, the probe signal reveals only the expectation value of the time-averaged measurement operator $\overline{\langle \sigma_z (t)\rangle} = 0 $, leading to the erasure of any information contained in the probe about the qubit state and thus canceling the measurement back-action on the qubit. In the language of the Stern-Gerlach experiment, the fast rotation of the spin allows the electron to fly through the measurement apparatus in a straight line without experiencing a force. Thus we call this effect quantum rifling, in analogy to the rifling of bullets, which stabilizes the trajectory of the projectile (see Figure  \ref{fig:sterngerlach}b). We then investigate the driving threshold to achieve rifling by measuring the Rabi decay rate of a probed qubit for different probe field amplitudes. Finally, using tomographic reconstruction of the qubit's state, we demonstrate read out multiplexing of two qubits coupled to the same measurement apparatus: quantum rifling is used to suppress the measurement back-action on one of the qubits on demand, while still extracting full information about the state of another qubit. 

Our system consists of two superconducting transmon qubits\cite{Koch2007} coupled dispersively to a microwave resonator\cite{blais2004cavity}(see Sec. I of Supplementary Material\cite{supplementary} for details).  
The driven stationary microwave mode passing through the resonator acts as our measurement probe: its interaction with the qubit leads to a qubit state-dependent dispersive shift of the resonator frequency. Rifling of the qubit state is achieved by applying a resonant Rabi drive to a charge line coupled directly to the qubit. For weak Rabi driving, the resonator transmission measurement returns two peaks weighted by the corresponding populations of the ground and excited states of the qubit (see Figure  \ref{fig:sterngerlach}c). When the driving strength reaches a threshold,  the transmission spectrum yields a single central peak analogous to a straight flight-path for the spin in the Stern-Gerlach apparatus.

%\section{Results}

\begin{figure}
    \centering
    \includegraphics[scale=0.5]{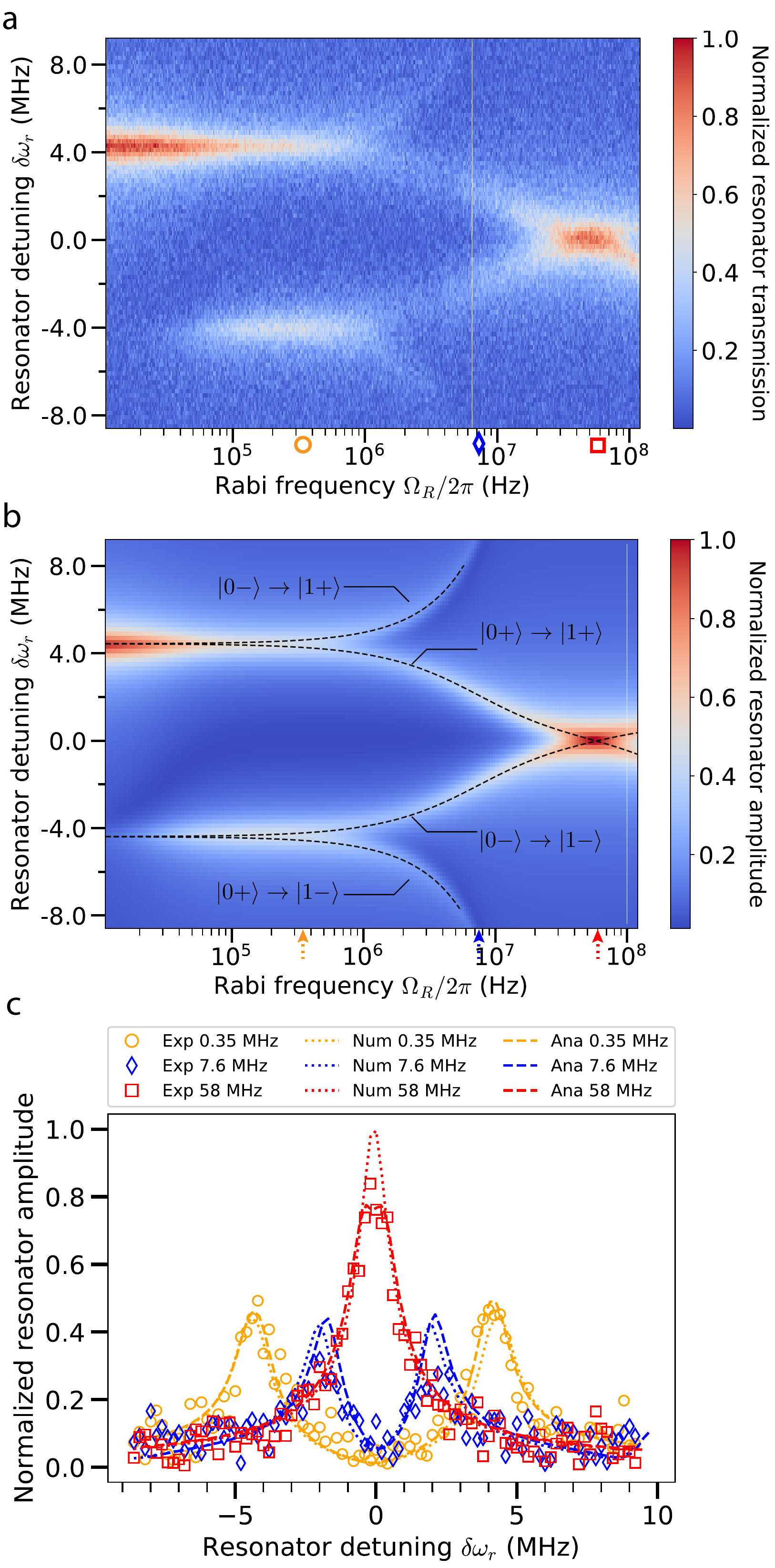}
    \caption{
    \textbf{Resonator spectroscopy vs Rabi drive strength. a,} Continuous wave spectroscopy of the resonator transmission vs. Rabi frequency. The resonator is probed by a mean photon number  of $\langle n_{photon}\rangle\approx 0.13$. The symbols below the plot show the qubit drive strength at which linecuts are shown in (c). \textbf{b,} Steady-state numerical simulation of Eq.(\ref{Master equation}) extended to the three-level transmon model for the observable $\vert a\rangle$ plotted as a function of $\delta\omega_r/2\pi$ and $\Omega_R/2\pi$, assuming a steady-state drive and using experimental parameters quoted in the main text. Dashed lines show the transitions in frequency between the $\vert 0\pm\rangle$ and $\vert 1\pm\rangle$ states.The arrows under the plot show the frequencies for the line cuts in (c). \textbf{c, }Comparison of  experimental data and  numerical simulations presented in (a-b), as well as the analytical solution for a two-level qubit at three values of the Rabi drive strength. All data have been normalized such that the maximum amplitude without Rabi drive is 1.}
    \label{fig:twinpeaks}
\end{figure}

%\subsection{Effects of rifling on the resonator}
Figure ~\ref{fig:twinpeaks}a shows the transmission spectroscopy of the resonator when Qubit 1 is continuously driven on resonance. Varying the qubit drive power we identify different characteristic regimes of the measurement. For low drive power the qubit remains in its ground state and only a single transmission peak is visible at $\omega_r+\chi$. As the Rabi drive becomes sufficient to excite the qubit, a second peak appears at $\omega_r-\chi$ in the cavity spectrum. The two peaks reach equal height when the qubit is saturated by the drive and the populations of the qubit in its ground and excited state becomes equal.

As $\Omega_R$ increases the two cavity peaks split, and the outer diverging peaks vanish with increasing drive. The inner peaks converge to a single peak at the average frequency $\omega_r$, which we identify as the onset of the quantum rifling regime. When the Rabi drive becomes comparable to the transmon anharmonicity,  the central cavity peak splits again as we populate higher levels, setting an upper bound on the rifling power (see Supplementary information Sec. II.).

The system can be described by the Jaynes-Cummings Hamiltonian in the dispersive regime transformed into a doubly rotating frame at both the qubit and probe drive frequencies $\omega_q,\omega_p$ respectively:
\begin{equation}
    H/\hbar= \delta\omega_r a^{\dagger}a  +\chi a^{\dagger}a\sigma_z
    +\frac{1}{2}\epsilon_d (a^{\dagger}+a)
    +\frac{1}{2}\Omega_R\sigma_x,
    \label{eq:H}
\end{equation}
where $\delta\omega_r=\omega_r-\omega_p$, $a^{\dagger}$ and $a$ are creation and annihilation operators of resonator excitation modes, $\sigma_{x/z}$ are Pauli matrices acting on the qubit, $\epsilon_d$ is the probe amplitude, and we assume resonant driving with the qubit's ground to excited state transition frequency, $\omega_q = \omega_{ge}$.
Including decoherence and losses for such a system\cite{Ithier2005}, the full time-evolution is described by the master equation 
\begin{align}\label{Master equation}
    \dot \rho =  -\frac{i}{\hbar} \left[ H_, \rho \right] + \kappa\mathcal D[a]\rho
        + \gamma_{\downarrow} \mathcal D[\sigma_-]\rho + \frac12 \gamma_\varphi \mathcal D[\sigma_z] \rho,
\end{align}
where $\mathcal D[o]\rho = o\rho o^{\dagger} -\frac12(o^{\dagger} o \rho + \rho o^{\dagger} o)$ , $\kappa/(2\pi) = 0.95~$MHz is the cavity decay rate and  $1/\gamma_\downarrow$, $1/\gamma_\varphi$  are the qubit relaxation and pure dephasing times, respectively.

We compute both numerically\cite{johansson2013} and analytically  the steady-state of the resonator probe amplitude $\langle a \rangle$ by solving Eq.~\eqref{Master equation} (as well as its extension to a three-level transmon) in the low photon number limit $\langle n_{photon}\rangle\approx 0$, plotting the result in Figure  \ref{fig:twinpeaks}b-c. 
The additional splitting of the central resonance peak around $\Omega_R/(2\pi)\approx 100$ MHz is well accounted for by the multi-level model.

When truncated to the single cavity photon subspace, the diagonalization of the Hamiltonian~\eqref{eq:H} for $\epsilon_d=0$   leads to four eigenstates in the dressed-state picture\cite{golter2014,Laucht2016}, each a superposition of both atom and resonator states. Note that these dressed states are of quantum origin leading to entanglement between the cavity field and the qubit, as opposed to e.g. the Mollow-tripet, where the qubit is dressed classically. The unnormalized expressions are:
\begin{align}
\label{eq:eigenstates}
    \vert 0\pm\rangle & \sim \vert g,0\rangle \pm \vert e,0\rangle \,,\nonumber\\
    \vert 1\pm\rangle & \sim \frac{2\chi\pm \sqrt{4\chi^2+\Omega_R^2}}{\Omega_R}\vert g,1\rangle + \vert e,1\rangle \,,
\end{align}
where $\vert{g,e}\rangle$ are the ground- and excited states of the qubit and $\vert{n}\rangle$ are resonator states with $n$ photons.
The corresponding eigenenergies in the rotating frame at $\delta\omega_r=0$ can be found as $E_{0\pm} = \pm\frac{1}{2}\hbar\Omega_R$ and $E_{1\pm} = \pm\frac{1}{2}\hbar \sqrt{4\chi^2+\Omega_R^2}$. 

The four cavity transmission peaks correspond to the four transitions between eigenstates with differing photon parity (see Figure \ref{fig:twinpeaks}b dashed lines). In the limit of  $\Omega_R\ll\kappa$, $E_{0-}\approx E_{0+}$ yielding two degenerate transitions at frequencies $\omega_r \pm \chi$. The splitting of the peaks is proportional to $\Omega_R$, which becomes visible when their splitting exceeds the linewidth $\Omega_R/2\pi>\kappa/2\pi \simeq 1$~MHz. 
In the opposite limit of strong driving $\Omega_R \gg \chi$, thus $E_0\approx E_1$. The matrix elements corresponding to the outer transitions at $E_{0\mp} \rightarrow E_{1\pm}$ vanish for large $\Omega_R$, making the outer peaks disappear. The inner transitions $E_{0\pm} \rightarrow E_{1\pm}$ converge in energy, merging into the central cavity peak once their separation is too small to still resolve two peaks $\vert E_1-E_0\vert/\hbar<\kappa$, leading to the condition for the quantum rifling regime  $\Omega_R > \Omega_C = \chi^2/\kappa\approx 2\pi\times 16$~MHz. 
%The emergence of a single peak can also be understood by observing that in the strong driving limit the dressed eigenstates are equal superposition of qubit ground and excited states, leading to a vanishing dispersive shift of the cavity and inability to obtain any information about the qubit.
For larger cavity probe powers, the evolution of the transmission peaks remains qualitatively the same, with less pronounced side peaks and the critical drive amplitude $\Omega_C$ shifting to higher drive frequencies (see Supplementary information Sec III. for details).

We have also verified that the effect of two resonator transmission peaks merging into one  can also be induced by fast incoherent qubit dynamics (see Supplementary information Sec. IV.). As for a qubit interacting with a heat bath at different temperatures such that $\langle \sigma_z\rangle\neq 0$ in the steady state, simulations show that the position in frequency of the central peak shifts proportionally to the asymmetry in population between the ground and excited state (see Supplementary information Sec. II. B1).

\begin{figure}[ht!]
   \centering
    \includegraphics[width=0.5\textwidth]{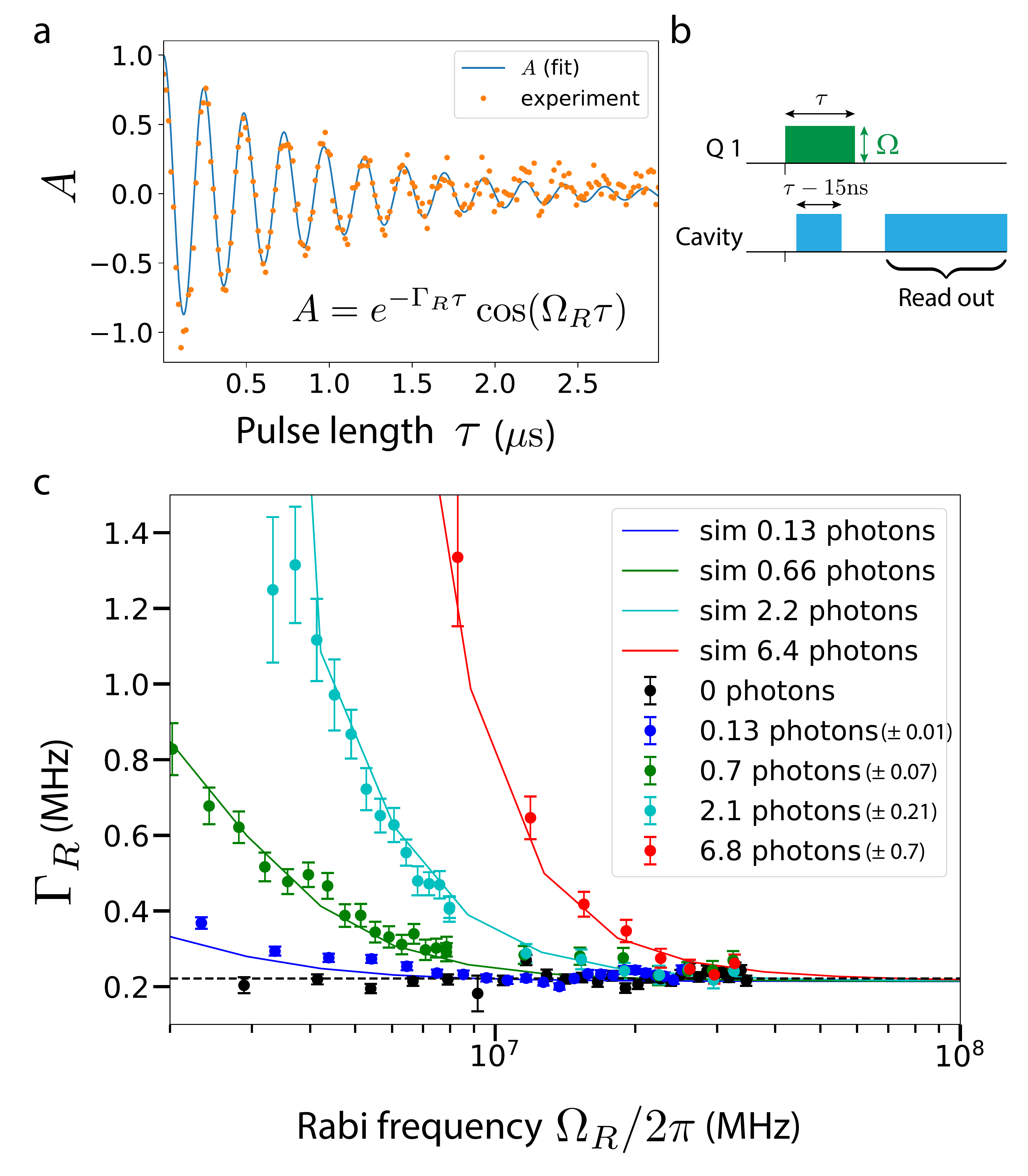}
    \caption{
    \textbf{Rabi decay rates vs. Rabi frequency $\Omega_R$ for different measurement strengths. a,} Example of a rifled Rabi measurement ($\Omega_R/2\pi = 4.1$ MHz, $\langle n_{photon}\rangle\approx 2.1$) with overlaying numerical fit of the equation shown in the insert. \textbf{b, }Pulse protocol of the rifled Rabi measurement. The cavity pulse starts 10 ns after the qubit drive and stops 15 ns before its end. The actual read out pulse starts 5 ns after the qubit drive ends. \textbf{c,} Measured and simulated rifled Rabi and standard Rabi decay rates. For a fixed probe power, the Rabi decay rate is measured for increasing Rabi drives following the protocol presented in (b). The values for $\langle n_{photon}\rangle$ are measured independently via the ac-Stark shit. For the simulations, the average number of photons is extracted by normalizing to the fitted curve for the lowest measurement strength, which is assumed to match the measurement.
    }
    \label{fig:rabitimes}
\end{figure}

%\subsection{Qubit coherence during rifling}
We examine the back-action of the detector on the qubit by measuring the relaxation rate of Rabi oscillations with and without a simultaneous resonator tone. The qubit relaxes due to measurement back-action and through intrinsic losses to the environment. For low Rabi frequencies, the back-action dominates this relaxation. However, with increasing Rabi frequency the detector's ability to distinguish between the qubit eigenstates diminishes thus  reducing the measurement back-action, reaching zero in the single peak regime.
The results are presented in Figure \ref{fig:rabitimes}, varying the Rabi frequency for each specific probe power.

For a Rabi drive $\Omega_R \ll \Omega_C$, the Rabi coherence time $T_R = \Gamma_R^{-1}$ is degraded by the measurement: the probe extracts information from the qubit leading to its dephasing. For strong Rabi drive $\Omega_R \gg \Omega_C$, the qubit coherence times are comparable to the standard Rabi decay time measured without applying a simultaneous cavity pulse. Driving the cavity with more photons leads to a stronger suppression of the coherence time for small Rabi frequencies, consistent with the measurement rate being proportional to the cavity’s photon number population. The threshold Rabi drive frequency $\Omega_C$ at which $T_R$ converges to the standard Rabi coherence time of the qubit is, however, independent of the cavity tone strength for low photon population ($\langle n_{photon}\rangle \ll 1$). This is in agreement with the observed threshold drive required for the emergence of a single cavity peak in continuous wave spectroscopy (see Figure  \ref{fig:twinpeaks} and Supplementary information Sec. III.).

We also numerically simulate such rifled Rabi oscillations for Qubit 1 in the time domain and plot their decay rate in Figure \ref{fig:rabitimes}c.
Note that only a single fit parameter was used to scale the curve with lowest drive strength.
%Note that no free parameters are involved in fitting, besides an overall scaling factor used to extract the average photon number when compared to the curve with lowest drive strength.

%\subsection{Multiplexing two qubits using rifling}
As it suppresses the measurement back-action, rifling allows measurements on other qubits coupled to the same detector, while keeping the rifled qubit in superposition. 

\begin{figure}[h!]
    \centering
    \includegraphics[width=\linewidth]{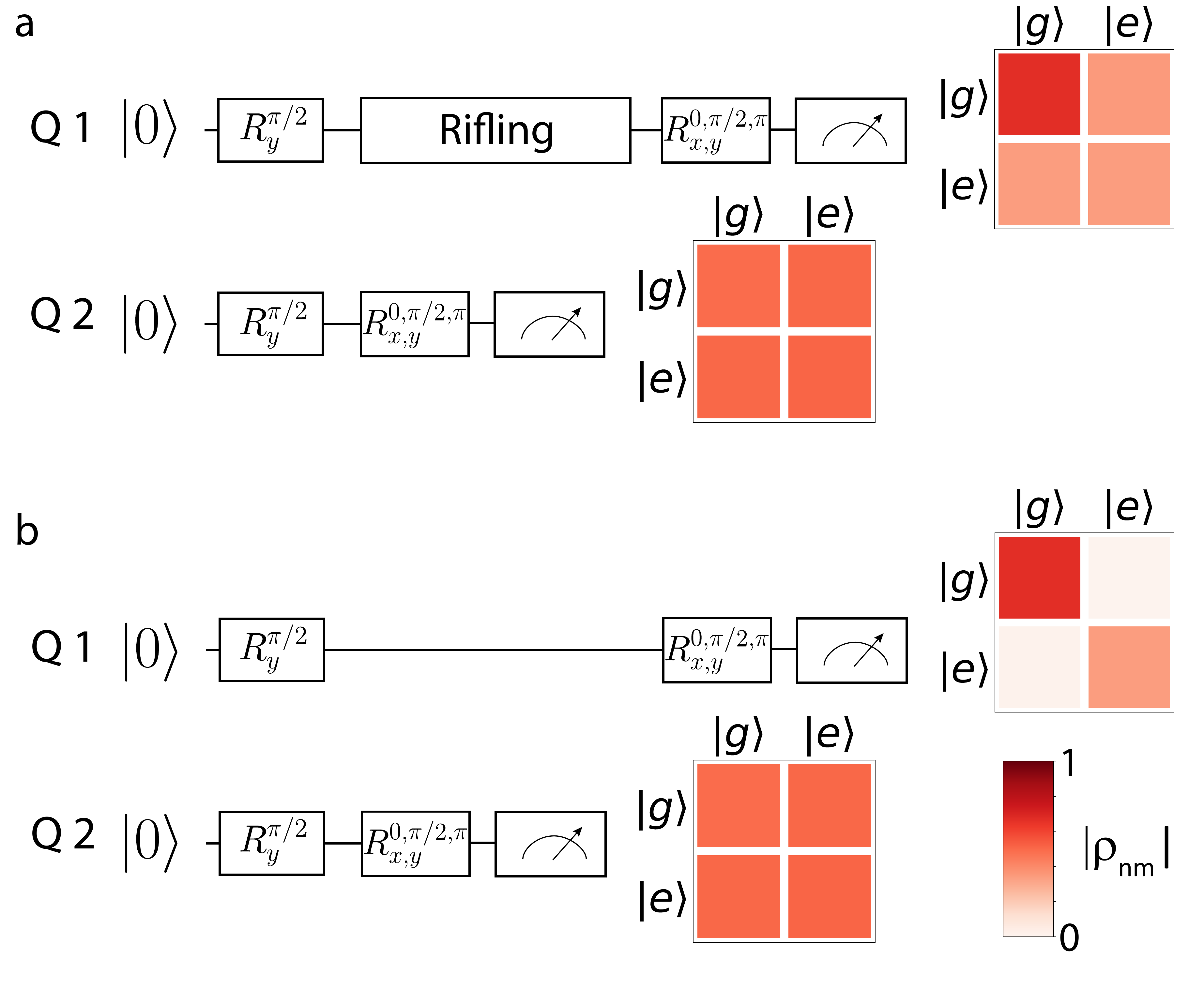}
    \caption{\textbf{Sequential measurement of two qubits coupled to the same resonator with and without rifling. a,} Gate protocol and measured single qubit density matrices. First, both qubits are brought to full superposition by applying a $R_y^{\pi/2}$ rotation. Then Qubit 1 is rifled by applying a strong coherent Rabi drive for 1142 ns, while Qubit 2 is read out via tomography pulses followed by a cavity read out pulse, which ends before the Rabi drive pulse on Qubit 1. Lastly, Qubit 2 is read out.\textbf{b,}  Same as in (a), but Qubit 1 is left idle for 1142 ns instead of rifling.
    }
    \label{fig:2qubits}
\end{figure}

To demonstrate this multiplexing capability, we perform a two-qubit algorithm with  both qubits coupled to the same readout resonator (see Figure \ref{fig:2qubits}). First we prepare a superposition of the four basis states by applying a $R_y^{\pi/2}$ rotation to both qubits. We then rifle Qubit 1 for 1142 ns, while performing tomography on Qubit 2, followed by tomography on Qubit 1. We measure the density matrix of Qubit 1 with 92.8\% fidelity (corrected for qubit decoherence), confirming the qubit  remained in superposition following the first read out (Figure \ref{fig:2qubits}a).
Conversely, omitting the rifling pulse (Figure \ref{fig:2qubits}b) leads to vanishing non-diagonal terms in the density matrix and therefore to the collapse of coherence of Qubit 1, induced by the read out of Qubit 2.  

It is worth noting that following the initial $R_y^{\pi/2}$ pulse, the qubit can be rotated around two different axes of the Bloch sphere: either around the x-axis, inducing full rotations around the Bloch sphere, or around the y-axis, effectively spin-locking the qubit\cite{Yan2013b}. We report that in both cases the coherence is preserved (see Supplementary information Sec. V.). 
We also performed rifling of Qubit 2 while extracting information from Qubit 1 with similar results (see Supplementary information Sec. VI.).

Other protocols to suppress measurement back-action on qubits\cite{Suh2014,Touzard,Peronnin2019} have been recently realized; contrary to quantum rifling, however, these methods require additional time-dependent pump drives beyond the Rabi drive and thus offer a less practical implementation.
Interestingly, the merging of the two cavity peaks with increasing qubit modulation is reminiscent of motional averaging of a linewidth of a molecule\cite{Li2013}. 
In analogy, one could view our experiment as motional narrowing of the resonator, where the qubit acts as noisy environment shifting the resonator's resonance frequency. 
Likewise, the absence of measurement dephasing for a fast rotating qubit bears similarities with dynamical decoupling schemes\cite{Bylander2011}, where noise during successive periods of free evolution interferes destructively at specific moments in time. 
In contrast, quantum rifling allows controlled decoupling of individual qubits from their measurement apparatus for all times during rifling. In addition, quantum rifling could be used to render eigenstates of a system indistinguishable and thus avoid the collapse of superposition if one wishes only to distinguish between larger submanifolds of the system. For example, if one tries to measure whether a qutrit is in its ground state or not without destroying the coherence between its first and second excited states, one can achieve this by driving the transition between the two excited states during measurement.

%\section{Conclusion}
In conclusion, our results reveal an intuitive picture for the regime of a strongly driven, continuously measured qubit, where the Rabi frequency exceeds both the measurement rate and the meter bandwidth. 
In this regime, the resonator photons are not able to extract information about the qubit's state, leading to a time-averaged population measurement  of the qubit and importantly imposing no back-action. 
We have also demonstrated that a strong Rabi drive can be utilized as an experimental knob to tune the measurement back-action between qubit and probe, without affecting the  ability to measure other qubits probed simultaneously by the same field.
This capability allows for many qubits to be connected to an individual detector, thus facilitating scalability in architectures where connecting individual detectors to every qubit might be technically challenging\cite{Heinsoo2018}.

%increasing the quantum volume\cite{cross2018validating} of architectures previously limited to pairwise qubit coupling\cite{Heinsoo2018}, and thus  facilitating the scaling of future quantum processors.

\begin{acknowledgements}

We thank Benjamin Huard, Gerard Milburn, Tom Stace and Sergey N. Shevchenko for fruitful discussions.
This research was supported by the Australian Research Council Centre of Excellence for Engineered Quantum Systems (EQUS, CE170100009), the ARC Future Fellowship FT140100338 and by the Swiss National Science Foundation through the NCCR QSIT.

\end{acknowledgements}
 
%\printbibliography
\bibliography{references.bib}

%\bibliography{refs}% Produces the bibliography via BibTeX.

\end{document}

\title{Supplementary information: Quantum rifling - protecting a qubit from measurement back-action}

\author{Daniel Szombati}
\affiliation{ARC Centre of Excellence for Engineered Quantum Systems, Queensland 4072, Australia}
\affiliation{School of Mathematics and Physics, University of Queensland, St Lucia, Queensland 4072, Australia}

\author{Alejandro Gomez Frieiro}
\affiliation{ARC Centre of Excellence for Engineered Quantum Systems, Queensland 4072, Australia}
\affiliation{School of Mathematics and Physics, University of Queensland, St Lucia, Queensland 4072, Australia}

\author{Clemens M\"uller}
\affiliation{IBM Research Z\"urich, 8803 R\"uschlikon, Switzerland}
%\affiliation{Institute for Theoretical Physics, ETH Zürich, Switzerland}

\author{Tyler Jones}
\affiliation{ARC Centre of Excellence for Engineered Quantum Systems, Queensland 4072, Australia}
\affiliation{School of Mathematics and Physics, University of Queensland, St Lucia, Queensland 4072, Australia}

\author{Markus Jerger}
\affiliation{ARC Centre of Excellence for Engineered Quantum Systems, Queensland 4072, Australia}
\affiliation{School of Mathematics and Physics, University of Queensland, St Lucia, Queensland 4072, Australia}

\author{Arkady Fedorov}
\affiliation{ARC Centre of Excellence for Engineered Quantum Systems, Queensland 4072, Australia}
\affiliation{School of Mathematics and Physics, University of Queensland, St Lucia, Queensland 4072, Australia}

\date{\today}
\onecolumngrid
\maketitle
\tableofcontents

\newpage

\section{Sample characteristics and electronic measurement setup}

\begin{figure}[h!]
    \centering
    \includegraphics[width=0.8\linewidth]{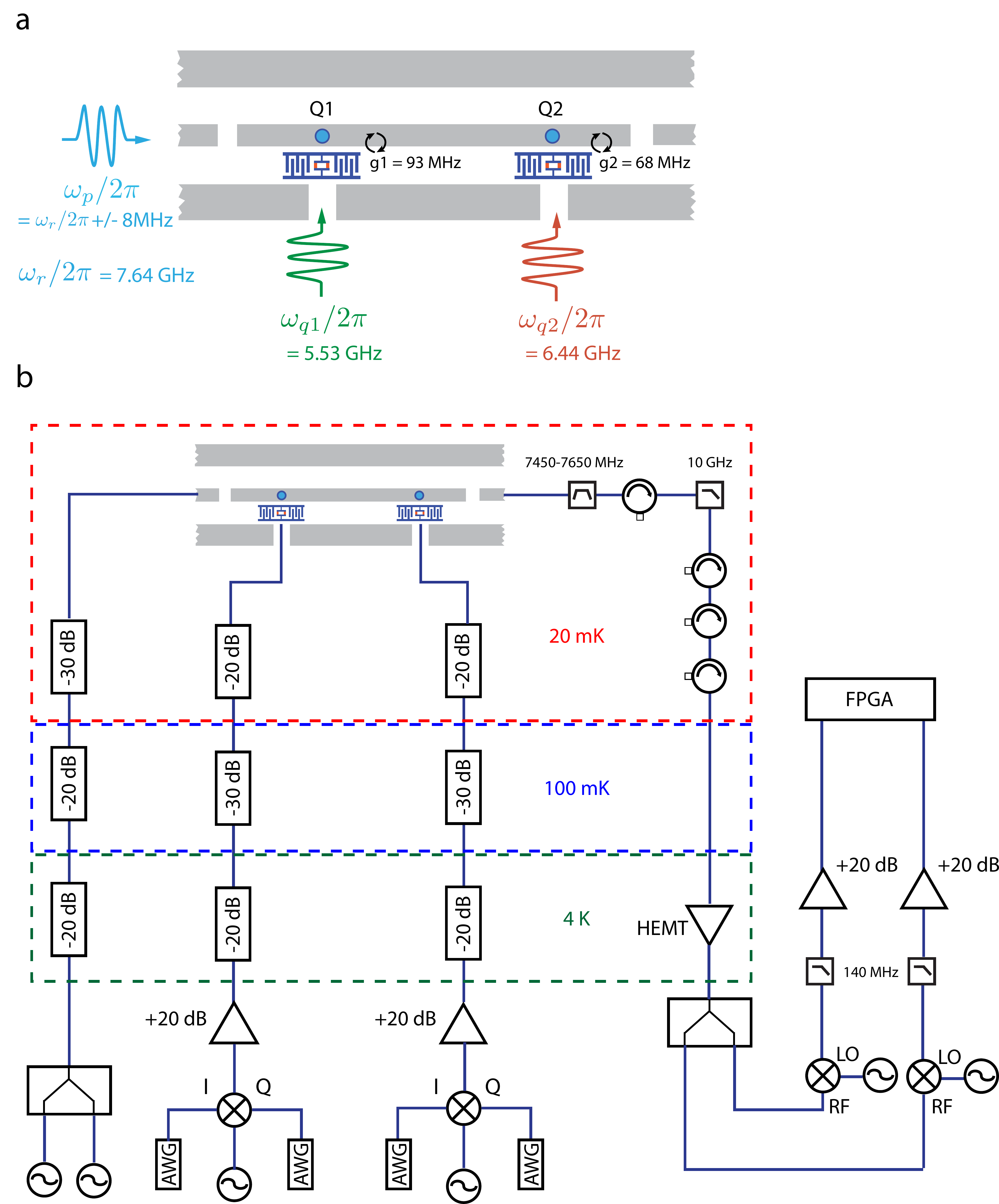}
    \caption{\textbf{Schematics of the electronic measurement setup. a,} Sketch of the two qubits coupled to the resonator and the driving tones used in the experiment. \textbf{b,} Sketch of the room temperature and cold electronics setup. For details of each individual component refer to the Supplementary text.}
    \label{fig:setup}
\end{figure}

\subsubsection*{Sample characteristics}

The sample consists of four 2D transmon qubits pair-wise coupled to four superconducting co-planar waveguide in a circuit-QED arrangement. Only two qubits and one resonator were used for the experiment. The sample substrate is sapphire and the junctions were fabricated using a double-angle aluminum shadow evaporation technique. The resonator's central resonance frequency $\omega_r/2\pi = 7.641$ GHz and its decay rate is $\kappa/(2\pi)=0.95$ MHz$(\pm 60.1$ kHz). Qubit1 is coupled to the resonator with a rate $g_1 = 93.1$ MHz, resulting in a dispersive shift $\chi/(2\pi)= 4.1$ MHz$(\pm 79.9$ kHz), has maximum ground to excited state transition frequency of $\nu^{max}_{ge}=5.53$ GHz and the relaxation times $T_1= 5.1$ $\mu$s$(\pm 0.04$ $\mu$s) $T_2^\text{{Ramsey}}=5.91$ $\mu$s. For Qubit2 $g_1 = 68.4$ MHz, $\chi/(2\pi)=3.9$ MHz$(\pm 54.0$ kHz),  $\nu^{max}_{ge}=6.44$ GHz and relaxation times $T_1=1.87$ $\mu$s$(\pm 0.05$ $\mu$s) $T_2^\text{{Ramsey}}=$3.39 $\mu$s. Both qubits are operated at their maximum transition frequencies.

Measurement fidelities were calculated using the equation $F =[\mathrm{Tr}\hspace{1pt}(\sqrt{\sqrt{\rho}\rho_{exp}\sqrt{\rho}})]^{2}$.

\subsubsection*{Brief description of the setup}
The measurements were performed at the University of Queensland, in an Oxford Cryogenics dilution refrigerator(DR) with base temperature of 20 mK. Schematics of the room and cold temperature electronics are shown in Fig. \ref{fig:setup}. The cavity probe tones were applied by combining the signals from a Rohde \& Schwarz SMB-100 and an Agilent E8257D microwave generator via a Mini-Circuits ZX 10-2-98-S+ splitter, which is then sent via an attenuated line in the dilution fridge to the cavity. The qubit drive pulses (for both qubits) are generated mixing the continuous signal from a Rohde \& Schwarz SMB-100 microwave generator with pulses generated via a Tektronix AWG5014C AWG, mixed using a Marki IQ-0307 MXP mixer. The combined signal is attenuated inside the DR. The output of the cavity is filtered through a Mini-Circuits VBF-78500+ bandpass filter followed by a Raditek RADC 4-8 Cryo circulator, then a Mini-Circuits VLFX-1050 10 GHz low pass filter and another three Raditek RADC 4-8 Cryo circulators. This output signal is then amplified at the 4K stage using a CMT CIT CRYO 4-12A HEMT cryogenic low noise amplifier. At room temperature, the output is split using a Mini-Circuits ZX 10-2-98-S+ splitter, such that both cavity input sources have their own down conversion source. Both branches are then down converted to 25 MHz by mixing the signal with a continuous tone from a Rohde \& Schwarz SGS 100-A microwave source using Mini-Circuits ZMX-10 G+ mixers. The down converted signal is then filtered by Mini-Circuits SLP-150+ low pass filters and amplified by  Mini-Circuits ZFL-500LN+ amplifiers. Finally the two channels from the cavity are digitized by a M4i.4450-x8 digitizer and read by a Xilinx Virtex 4 FPGA FPGA card, digitally down converted and both quadratures are recorded. For continuous wave steady state experiments, the same setup was used but with the mixers and AWG's omitted, leaving the microwave source connected to the qubit drive line. The setup generating the incoherent noise on the qubit is detailed in section\ref{sec:incoherent_drive} of this Supplementary.
\newpage

\section{Theoretical calculations}

\subsection{Multi-level model for transmon}

For increased accuracy, and to accurately predict the behavior of the system at very high Rabi powers, we model the transmon qubit as a multi-level system. Only the qubit parts of the master equation will change compared to equations in the main text. For the qubit Hamiltonian we then write
\begin{align}
	H_{q} = \sum \omega_{k} \sigma_{k,k}
\end{align}
where we defined the qubit operators $\sigma_{k,l} = \ket k \bra l$ with the qubit eigenstates $\ket k$ and the qubit eigenenergies $\omega_{k}$. 
We additionally assume that $\omega_{k} = k \omega_{q} - \sum_{n=1}^{k} n \alpha_{q}$, which is a common approximation for a weakly anharmonic qubit like the transmon. 

The dispersive coupling term changes to
\begin{align}
	H_{disp} = \sum_{k} (\chi_{k} - \chi_{k-1}) \sigma_{k,k} a\hc a \,,
\end{align}
with the dispersive shifts $\chi_{k} = \frac{g_{k}^{2}}{\omega_{r} - \omega_{k+1,k}}$, where $g_{k} \approx \sqrt{k+1} g_{0}$ is the coupling strength between the qubit's 
$\ket k \leftrightarrow \ket{k+1}$ transition and the resonator and $\omega_{k+1,k} = \omega_{k+1} - \omega_{k} \approx \omega_{q} - k \alpha_{q}$ is the energy splitting of that qubit transition. 

The qubit dissipative contributions to the master equation change to
\begin{align}
	\mathcal L_{q} \rho = \sum_{k} \gamma_{k,\downarrow} \diss{\sigma_{k,k+1}}\rho + \sum_{k} \gamma_{k,\uparrow} \diss{\sigma_{k+1,k}}\rho 
		+ \frac12 \gamma_{\varphi} \diss{\sum_{k} \sigma_{k,k}}\rho
\end{align}
where, under the assumption of weak anharmonicity of the qubit, we again approximate $\gamma_{k} \approx k \gamma$. 

\subsection{Analytical solutions for the two-level model}

Calculating the equation of motion for the system observables obeying Eq. 3 of the main text, we find the coupled system of equations
\begin{align}
	\partial_{t} \mean{a} =& -\ii \l( \delta\omega_{r} \mean a +\frac12\epsilon_{d} + \chi \mean{\sigma_{z} a} \r) - \frac{\kappa}{2} \mean{a} 
		\label{eq:ResA}\\
	\partial_{t} \mean{a\hc a} =& \ii \frac{\epsilon_{d}}{2} \l(\mean a - \mean{a\hc}\r)  - \kappa \mean{a\hc a} \\
	\partial_{t} \mean{\sigma_{z}} =& \ii \Omega_{R} (\mean{\sigma_{+}} - \mean{\sigma_{-}}) 
		+ \gamma_{\downarrow} - \gamma_{\uparrow} - (\gamma_{\downarrow} + \gamma_{\uparrow}) \mean{\sigma_{z}} \\
	\partial_{t}\mean{\sigma_{-}} =& -\ii \l( \delta\omega_{q} \mean{\sigma_{-}} -2\chi \mean{a\hc a \sigma_{-}} +\frac12\Omega_{r}\mean{\sigma_{z}} \r) 
		-\gamma_{2} \mean{\sigma_{-}} \\
	\partial_{t} \mean{\sigma_{z} a} =& -\ii \Bigl( \delta\omega_{r} \mean{\sigma_{z} a} + \frac{\epsilon_{d}}{2} \mean{\sigma_{z}} + \chi \mean a  
		- \Omega_{R} ( \mean{\sigma_{+} a} - \mean{\sigma_{-} a}) \Bigr) 
		- (\gamma_{\uparrow} - \gamma_{\downarrow}) \mean a - ( \gamma_{\downarrow} + \gamma_{\uparrow} + \frac{\kappa}{2} ) \mean{\sigma_{z} a} \\
	\partial_{t}\mean{\sigma_{-}a} =& - \ii \Bigl( \delta\omega_{r} \mean{\sigma_{-} a} - \chi \mean{\sigma_{-} a} -2\chi \mean{\sigma_{-} a\hc a a}  
		+\frac{\epsilon_{d}}{2} \mean{\sigma_{-}}+\delta\omega_{q} \mean{\sigma_{-}a} +\frac{\Omega_{R}}{2} \mean{\sigma_{z}a} \Bigr) 
		- (\gamma_{2} + \frac{\kappa}{2}) \mean{\sigma_{-}a} \\
	\partial_{t}\mean{\sigma_{+}a} =& - \ii \Bigl( \delta\omega_{r} \mean{\sigma_{+} a} + \chi \mean{\sigma_{+} a} + 2\chi \mean{\sigma_{+} a\hc a a}  
		+\frac{\epsilon_{d}}{2} \mean{\sigma_{+}}-\delta\omega_{q} \mean{\sigma_{+}a} - \frac{\Omega_{R}}{2} \mean{\sigma_{z}a} \Bigr) 
		- (\gamma_{2} + \frac{\kappa}{2}) \mean{\sigma_{-}a}
\end{align}
which we can solve for their steady-states, i.e. for $\partial_{t}\mean {\op o} = 0$.
Here we have defined the total dephasing rate of the qubit $\gamma_{2} = \gamma_{\varphi} + \frac12 (\gamma_{\downarrow}+ \gamma_{\uparrow})$. 

The above set of equations is incomplete, and does in fact not close due to the unbounded nature of the harmonic oscillator operators. In the following we will assume factorization conditions for two different cases which allow for an analytic solution. 

\subsubsection{Analytic solution for incoherent driving, $\Omega_{R} = 0$, infinite temperature}

For no Rabi-drive, $\Omega_{R} = 0$, the nuber of relevant equations reduces significantly and we can find the solutions for the steady-state
\begin{align}
	\overline{\mean{\sigma_{z}}} =& \frac{\gamma_{\downarrow} - \gamma_{\uparrow}}{\gamma_{\downarrow} + \gamma_{\uparrow}} \quad \,, \quad
	\overline{\mean{\sigma_{z} a}} = \frac{\l( \chi -\ii (\gamma_{\downarrow} - \gamma_{\uparrow}) \r)\overline{\mean{a}} + \frac12 \epsilon_{D} \overline{\mean{\sigma_{z}}}}
		{ \delta\omega_{r} - \ii \gamma_{\Sigma}}  \nn\\
	\overline{\mean{a}} =& -\frac12 \frac{\epsilon_{d}\l( 1-\ii \frac{\chi(\gamma_{\downarrow}  - \gamma_{\uparrow})}{(\gamma_{\downarrow} + \gamma_{\uparrow})(\gamma_{\Sigma} + \ii \delta\omega_{r})} \r)}{\delta\omega_{r} 
		- \ii\l( \frac{\kappa}{2} + \frac{\chi (\chi + \ii (\gamma_{\downarrow} - \gamma_{\uparrow}))}{\gamma_{\Sigma} + \ii \delta\omega_{r}} \r)} \,,
\end{align}
where $\gamma_{\Sigma} = \frac{\kappa}{2} + \gamma_{\downarrow} + \gamma_{\uparrow}$.
In the special case of infinite temperature in the qubit bath, $\gamma_{\uparrow} = \gamma_{\downarrow} = \gamma$, this further reduces to
\begin{align}
	\overline{\mean{\sigma_{z}}}_{\infty} =& 0 \quad \,, \quad
	\overline{\mean{\sigma_{z} a}}_{\infty} = \frac{\chi}{ \delta\omega_{r} - \ii \gamma_{\Sigma}} \overline{\mean{a}} \nn\\
	\overline{\mean{a}}_{\infty} =& -\frac12 \frac{\epsilon_{d}}{\delta\omega_{r}  + \frac{\chi^{2}\delta\omega_{r}}{\delta\omega_{r}^{2} + \gamma_{\Sigma}^{2}} 
		- \ii\l( \frac{\kappa}{2} - \frac{\chi^{2}\gamma_{\Sigma}}{\delta\omega_{r}^{2} + \gamma_{\Sigma}^{2}} \r)} \,.
\end{align}

\begin{figure}[htb!]
    \centering
    \includegraphics[width=0.325\linewidth]{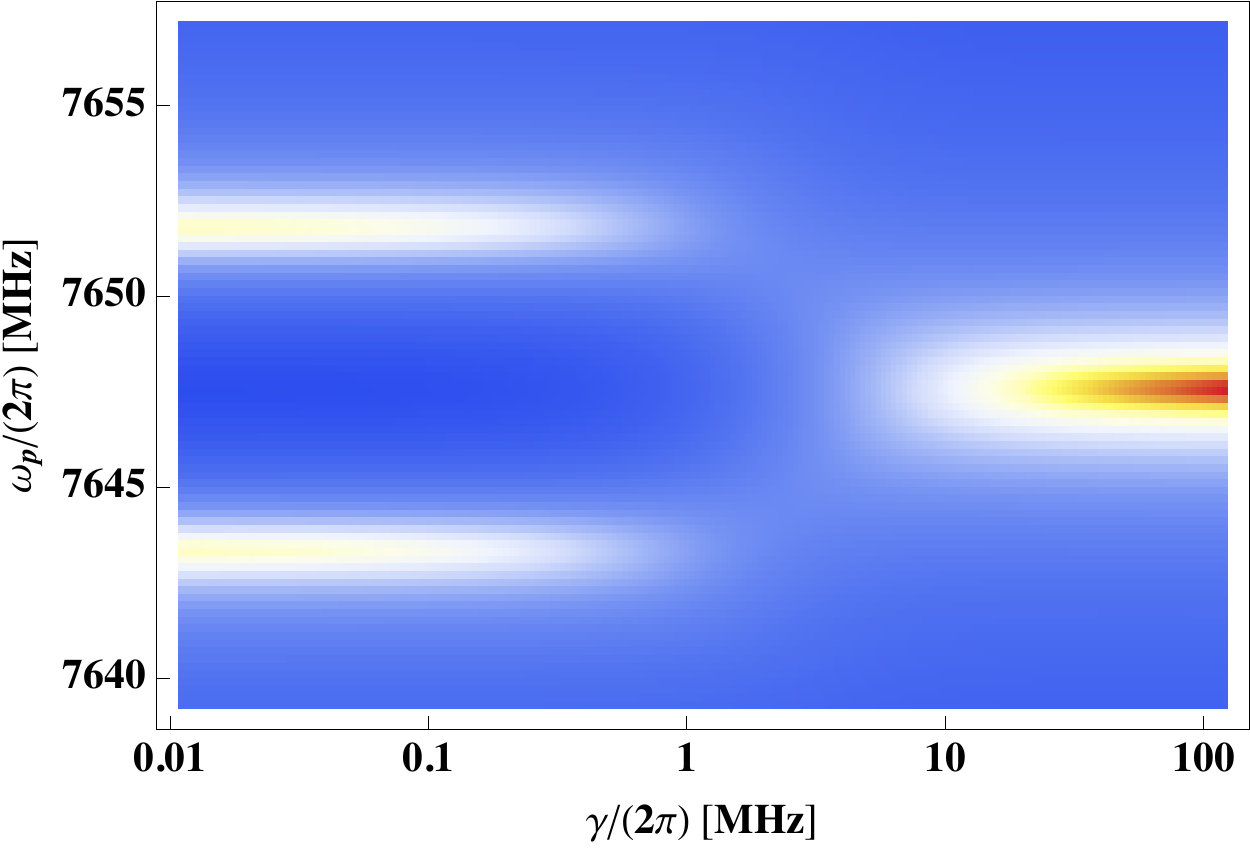}
    \includegraphics[width=0.325\linewidth]{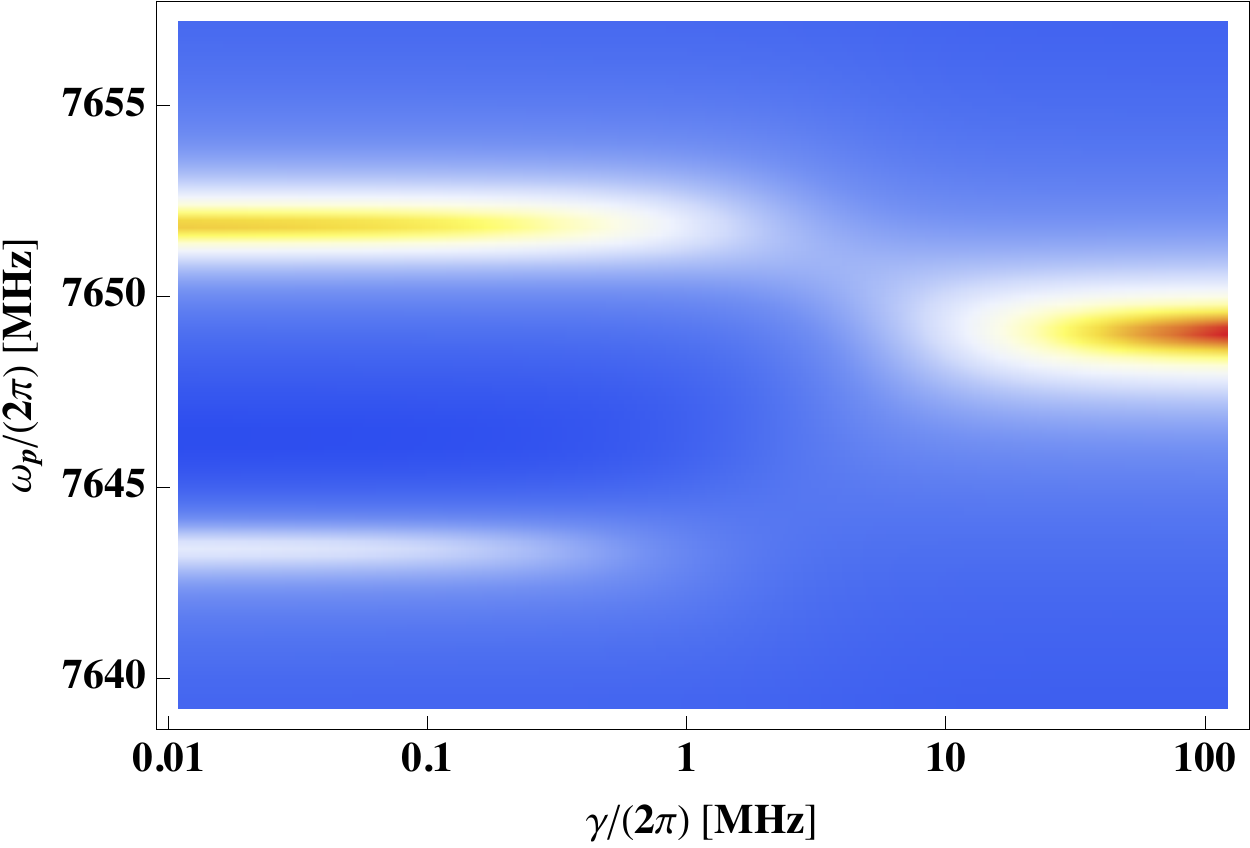}
    \includegraphics[width=0.325\linewidth]{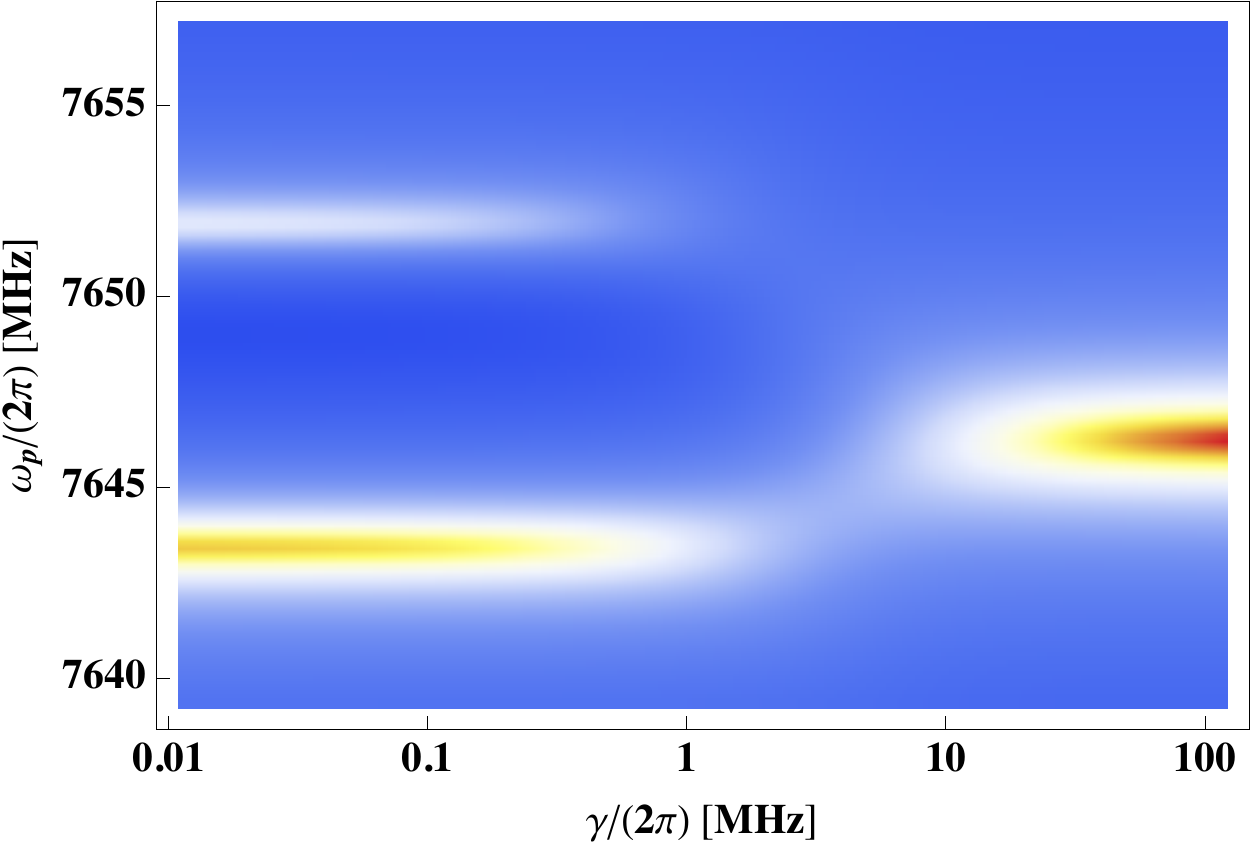}
    \caption{Analytical solution of the normalized steady-state resonator amplitude 
        for the incoherent driving case, for the situations of (left) infinite temperature, $\gamma_\uparrow = \gamma_\downarrow = \gamma$, (middle) high temperature such that $\gamma_\uparrow = \frac12 \gamma_\downarrow = \frac12\gamma$ and (right) negative temperature $\gamma_\downarrow = \frac 12 \gamma_\uparrow = \frac12\gamma$. 
        x-axis has the rate $\gamma$, all other parameters as in Fig 2 of the main text.
        }
    \label{fig:IncoherentDrive}
\end{figure}

\subsubsection{Analytical solution for arbitrary $\Omega_{R}\neq 0$, weak resonator drive}

To make the system of equations tractable, we will employ a truncation at correlation functions of at most second order, taking the approximation $\mean{\sigma_{-} a\hc a} \approx \mean{a\hc a}\mean{\sigma_{-}}$ and $\mean{\sigma_{-} a\hc a a} \approx \mean{a\hc a}\mean{\sigma_{-}a}$. 
Further we will assume that the resonator drive is weak, so that we can set $\mean{a\hc a} = 0$. Together these approximations lead to a linear and closed set of equations, which are amenable to an analytical solution. 

We then find the steady-state expressions
\begin{align}
	\overline{\mean{\sigma_{z}}} =& \frac{(\gamma_{\down} - \gamma_{\up})(\gamma_{2}^{2} + \delta\omega_{q}^{2})}{(\gamma_{\down} + \gamma_{\up}) (\gamma_{2}^{2}+\Omega_{R}^{2}) 
		+ 2\gamma_{2}\Omega_{R}^{2} } \\
	\overline{\mean{\sigma_{-}}} =& -\frac\ii2 \frac{(\gamma_{\down} - \gamma_{\up})(\gamma_{2} -\ii \delta\omega_{q}) \Omega_{R}}{ (\gamma_{\down} + \gamma_{\up}) (\gamma_{2}^{2}+\Omega_{R}^{2}) 
		+ 2\gamma_{2}\Omega_{R}^{2}  } \\
	\overline{\mean{\sigma_{-} a}} =& -\frac12 \frac{ \epsilon_{d} \overline{\mean{\sigma_{-}}} +\Omega_{r} \overline{\mean{\sigma_{z}a}} }{ (\delta\omega_{r} + \delta\omega_{q} - \chi) 
		-\ii \l( \gamma_{2} + \frac{\kappa}{2} \r) } \\
	\overline{\mean{\sigma_{+} a}} =& -\frac12 \frac{ \epsilon_{d} \overline{\mean{\sigma_{-}}} - \Omega_{r} \overline{\mean{\sigma_{z}a}} }{ (\delta\omega_{r} - \delta\omega_{q} + \chi) 
		-\ii \l( \gamma_{2} + \frac{\kappa}{2} \r) } \\
	\overline{\mean{\sigma_{z} a}} =& \frac{ \Omega_{R} ( \overline{\mean{\sigma_{+}a}} + \overline{\mean{\sigma_{-}a}} ) -\l( \chi -\ii (\gamma_{\up} - \gamma_{\down})  \r) \overline{\mean{a}} 
		- \frac{\epsilon_{d}}{2} \overline{\mean{\sigma_{z}}} 
		}{\delta\omega_{r} -\ii \l( \gamma_{\down} + \gamma_{\up} + \frac{\kappa}{2} \r)} \\
	\overline{\mean a} =& -\frac12 \frac{\epsilon_{d} + 2\chi \overline{\mean{\sigma_{z} a}}}{\delta\omega_{r} - \ii \frac{\kappa}{2}}\text{,}
	\label{eq:aSS}
\end{align}
from which one can find the full solutions for theses correlation functions in steady-state. 
The solution to \eqref{eq:aSS} is used for the analytical linecuts in Fig.2 of the main paper. We do not provide the full expression here as it is somewhat cumbersome.
%which are plotted in Fig.~\ref{fig:2} as function of $\delta\omega_{r}$ and $\Omega_{R}$ for parameters $\chi=1,\kappa=.5,\gamma_{\down}=0.1,\gamma_{\up} = \gamma_{\varphi}=0$.
As an illustrative limiting case, we may consider the infinite coherence case for the qubit $\gamma_{\downarrow}, \gamma_\uparrow, \gamma_\varphi\rightarrow0$, which leads to 
%As we do not expect the qubit decoherence to be significant for the qualitative effect, we can take the limit of $\gamma_{\downarrow}, \gamma_\uparrow, \gamma_\varphi\rightarrow0$ in the above expression, and get
\begin{align}
	\overline{\mean{a}} 
	%=& -\frac{\epsilon_{d}}{2} \l( \delta\omega_{r} -\ii \l(\frac{\kappa}{2} + \frac{\chi^{2}}{\ii\delta\omega_{r} 
	%		+\Omega_{R}^{2} \l( \frac{1}{ 2\ii(\delta\omega_{r} - \chi) + \kappa } + \frac{1}{ 2\ii (\delta\omega_{r} + \chi) + \kappa } \r) +\frac{\kappa}{2} }  \r) \r)^{-1} \nn\\
		=& -\frac{\epsilon_{d}}{2} \l( \delta\omega_{r} - \frac{\chi^{2}}{\delta\omega_{r} - \frac\ii2\kappa 
			-\ii \Omega_{R}^{2}\l( \frac{1}{ 2\ii(\delta\omega_{r} - \chi) + \kappa } + \frac{1}{ 2\ii (\delta\omega_{r} + \chi) + \kappa } \r)} - \frac\ii2\kappa \r)^{-1} \,,
\end{align}
which already reproduces all the features seen in Fig. 2a apart from the different visibility of the upper and lower lobes at low Rabi power (which stems from the thermal population of the qubit states under weak driving) and the additional splitting of the central peak at very high Rabi drive powers, induced by higher level excitations of the transmon.

\begin{figure}[htb!]
    \centering
    \includegraphics[width=0.5\linewidth]{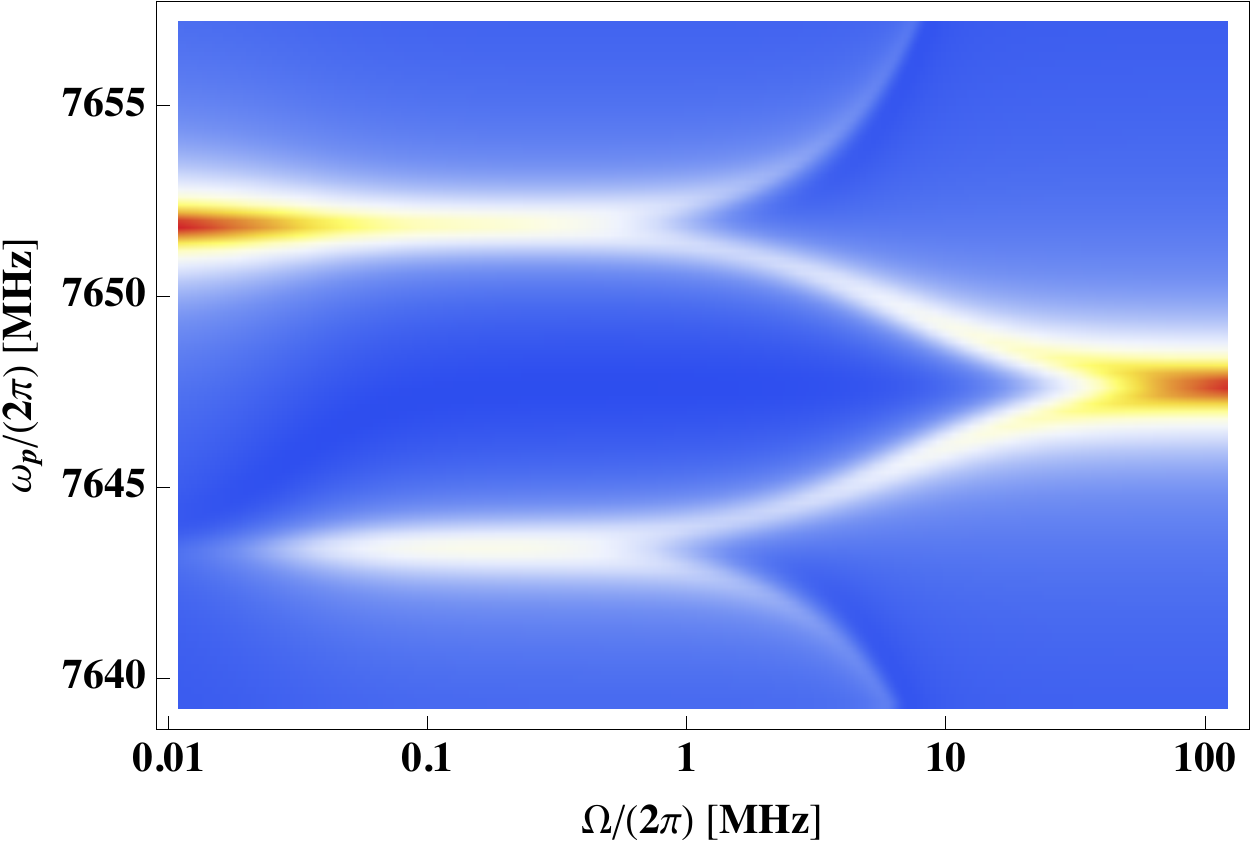}
    \caption{Analytical solution of the normalized steady-state resonator amplitude 
        for the coherent drive case in the two-level qubit model. 
        All parameters as in Fig.2 of the main text. 
        }
    \label{fig:IncoherentDrive}
\end{figure}

%\subsection{Analytical solution for detuned driving}

\subsection{Numerical calculations}\label{sec:Numerical}% - simulations with infinite temperature and asymmetric decay rates}

Some additional numerical simulations were performed using the python QuTiP library. The process to simulate the behaviour of the system is as follows. 
We first define an initial quantum state composed of a single qubit coupled to a resonator. We restrict the Hilbert space of the resonator to the first 30 levels. We also define sources of qubit ($\Gamma_1$) and cavity ($\kappa$) decay and qubit decoherence $\Gamma_{2}$ using the values presented in the main text for Qubit 1. These sources, characterized by some parameter $\Gamma_{i}$ are taken into account by collapse operators $C_{i}=\sqrt{\Gamma_{i}}A_{i}$, where $A_{1}$ is a operator related to $\Gamma_{i}$ that serves as a Lindbladian term in the master equation, as specified in the QuTIP documentation.

\subsubsection{Simulation of Rabi oscillations}

In order to simulate the time-dependent behaviour of the system, we take into account the Hamiltonian with two different drives (Eq. 2 of main text) and calculate the time evolution of the system with the function $qutip.mesolve$, obtaining the expectation value for the excited state population operator $| e \rangle\langle e |$ for a time up to 8000 ns. We perform this simulations for different values of the  Rabi frequency $\Omega_{R}$. In order to recreate the number of photons from the experiment, we calculate as well $\langle \hat{a}^{\dagger}\hat{a}\rangle$ with no qubit pulse ($\Omega_{R} = 0$), and use the appropriate values for $\epsilon_{d}$. Then, we fit the resulting Rabi oscillations to Eq. \ref{eq:Fit}. Simulation was done for $\kappa/2\pi = 0.00095$ kHz, $\Gamma_{1}=0.000195$ kHz and $\chi/2\pi = 0.0041$ kHz.

% \begin{equation}
% P_{e} (t)= Ae^{-\Gamma_{R} t}\cos(\Gamma_{R}t+\phi_{0})+B+Ce^{-\Gamma_{off}t}
% \label{eq:Fit}
% \end{equation}
% where $P_{e}$ is the excited state population, $A$ is the amplitude of the oscillation, $\Gamma_{R}$ is the Rabi frequency, $\phi_{0}$ is a time-independent initial phase of the oscillation, $B$ is a time-independent offset and $Ce^{-\Gamma_{off}t}$ is a time dependent offset, which is included to take into account the number of photons changing the dependence of Rabi oscillations. The results of these fits can be seen in Figure 3b in the main text.

\begin{equation}
P_{e} (t)= Ae^{-\Gamma_{R} t}\cos(\Omega_{R}t+\phi_{0})+B
\label{eq:Fit}
\end{equation}
where $P_{e}$ is the excited state population, $A$ is the amplitude of the oscillation, $\Omega_{R}$ is the Rabi frequency, $\phi_{0}$ is a time-independent initial phase of the oscillation, and $B$ is a time-independent offset. The results of these fits for the decay rate $\Gamma$ can be seen in Figure 3b in the main text for different number of photons.

\subsubsection{Estimating the phase-fluctation induced dephasing}

Following ITHIER et al. [PRB 2005, 72.13: 134519] we can estimate the echo dephasing due to phase fluctuations of the qubit as 
\begin{align}
%	T_{\varphi} = D_{\lambda} \sqrt{\frac{A}{2\pi} \log{\frac{A}{\omega_{IR}}}}
	T_{\varphi,E} = D_{\lambda} \sqrt{A \log{2} }
\end{align}
where we assume the qubit phase fluctuates according to
\begin{align}
	\Delta\phi(t) = D_{\lambda}\int_{t_{0}}^{t} dt'\: \delta\lambda(t')
\end{align}
where $\lambda$ is an external fluctuating parameter and 
$D_{\lambda} = \frac{\partial \omega_{q}}{\partial \lambda}$ encodes how changes in $\lambda$ influence the qubit level splitting $\omega_{q}$.
Additionally, we assume a spectral density of the fluctuations in $\lambda$ as
\begin{align}
	S_{\lambda} = \mean{\delta\lambda(t) \delta\lambda(0)} = \frac{A}{\abs{\omega}}
\end{align}
and we define the infrared cutoff frequency $\omega_{IR}$ as the inverse of the longest timescale present in experiments, such that the experiment cannot be influenced by any frequency below it. 
We also define the dephasing time as the time where the Echo decay function has decayed to a value of $1/\ee$ of its initial value.

Assuming $D_{\lambda} = 1$, or in other words direct frequency fluctuations of the qubit (or the pulse generator applying the Rabi pulse), we can calculate the dephasing contribution due to the phase noise of our source during the time comparable to the typical measurement time in circuit QED of $T_{\varphi} \sim 1~\mu$s.

Our microwave source has a phase drift of the order of $1^\circ$ every $10$ minutes. Assuming a diffusive random drift, this translates to a drift of the order of $10^{-8}$ rad after $1~\mu$s. This leads to a deviation from a pure state of magnitude $e^{-\frac{10^{-8}}{4}}\approx 1-\frac{10^{-8}}{4}$ (where the pure state is normalized to 1), in other words, the dephasing caused by the phase drift of the source is negligible.

\newpage

\section{Twin peaks vs Rabi coherence times plots}

In this section we present continuous-wave spectroscopy measurement of the resonator with simultaneous qubit driving, and compare the results with Rabi decay rate measurements with simultaneous cavity driving. We present four such data sets each measured with different steady-state average photon population in the resonator. 

\begin{figure}[h!]
    \centering
    \includegraphics[width=1.0\linewidth]{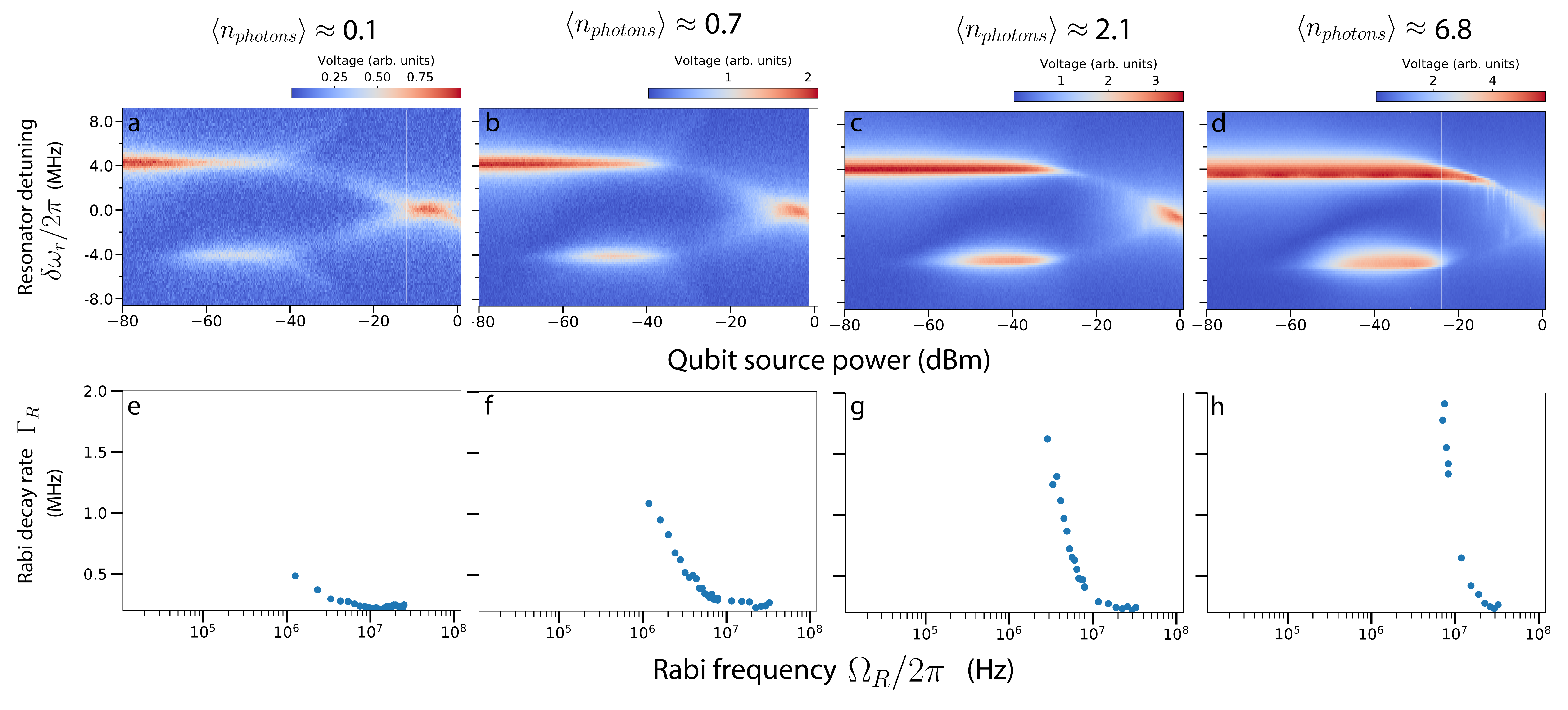}
    \caption{\textbf{Comparing quantum rifling of Qubit 1 for different measurement powers. a-d,}Resonator spectroscopy of Qubit 1 vs Rabi drive frequency for different resonator powers, as labelled above each panel. \textbf{e-h,} Rabi decay rate vs Rabi drive frequency of Qubit 1 for varying resonator drive strengths. The resonator is driven during the Rabi drive, with the same protocol as shown in Fig. 2a of the main text. The average photon number in the resonator during the first cavity pulse is shown above each panel. The second cavity pulse used for read out is kept fixed to 6.8 photons in the cavity. Data in panels (a) and (e)-(h) are also presented in the main text. The conversion between microwave power $A$ in dbm  and Rabi frequency $\Omega_R$ in Hz is $\Omega_R=11.2\times 10^{(A+19.85)/20+6}$ }
    \label{fig:twin_peaks_vs_rabi}
\end{figure}

Although the electronic setup is different for continuous-wave and time-domain measurements, we have converted the power applied from the qubit source to Rabi frequency seen by the qubit, by fitting the solutions of the dressed eigenstates (dashed line overlay in Figure 2b of the main text) on to the measured cavity peaks.  We present each panel \ref{fig:twin_peaks_vs_rabi}(a-d) such that the applied qubit power corresponds to the Rabi frequency in each panel below.

The aim is to reveal the correlation between the closing of the cavity peaks, and thus the effective coupling $\chi$, with the decrease in the Rabi decay rate $\Gamma_R$.

Panels \ref{fig:twin_peaks_vs_rabi}a,e-h is the same data also presented in the main text. 

\newpage

\section{Resonator spectroscopy with incoherent qubit drive}\label{sec:incoherent_drive}

\begin{figure}[h!]
    \centering
    \includegraphics[width=0.5\linewidth]{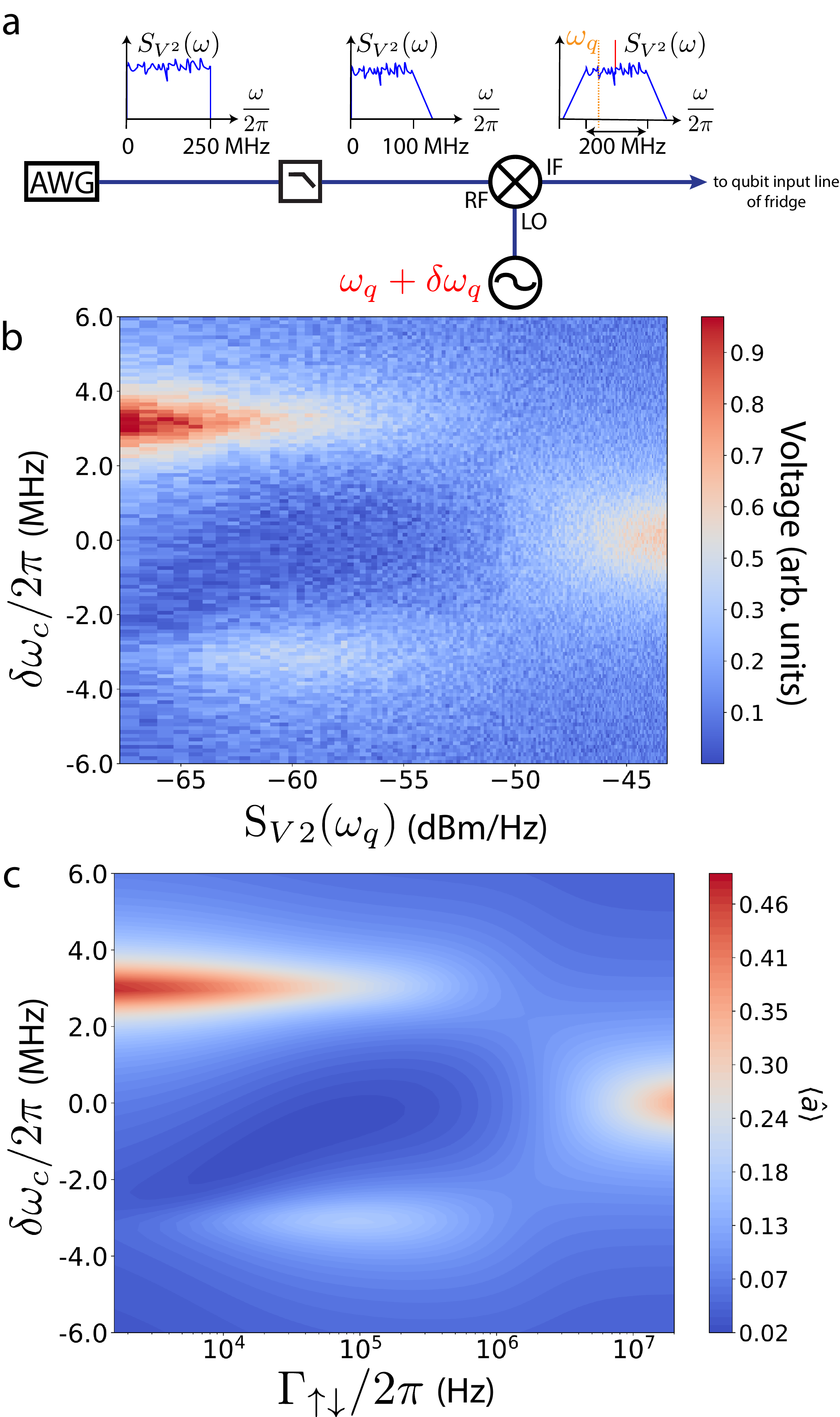}
    \caption{\textbf{Resonator transmission spectroscopy vs longitudinal qubit relaxation rate. a, }Room temperature electronics setup used to apply arbitrary stochastic noise to the qubit. The three inset graphs show the particular engineered microwave spectrum for different sections of the setup. \textbf{b, }Transmission spectroscopy measurement of the resonator vs resonator detuning and qubit relaxation rate. \textbf{c, }Numerical solution of the master equation using the Hamiltonian specified in Eq.3 of main text for $\Omega_{R} = 0$ vs $\omega_r$ and $\Gamma_{\uparrow\downarrow}$.}
    \label{fig:incoherent_drive}
\end{figure}

In this section we show measurements of the resonator spectroscopy where an incoherent drive on the qubit is present. The incoherent drive emulates a qubit environment in which the qubit's relaxation rates $\Gamma_{\uparrow}=\Gamma_{\downarrow}=\Gamma_{\uparrow\downarrow}/2$, and thus  $\Gamma_1$ depends on  $\Gamma_{\uparrow\downarrow} \propto S_{V^2}(\omega_{q})$, and is proportional to the incoherent drive power and can be experimentally controlled.  We achieve such a drive by mixing white noise centered at zero frequency with a high frequency local oscillator, as shown in Fig. \ref{fig:incoherent_drive}a. The measurement results are presented in  Fig. \ref{fig:incoherent_drive}b. This is to be compared with Fig. \ref{fig:incoherent_drive}c, where we plot the analytical solution to Eq. 3 of the main text, the Hamiltonian of a qubit coupled to an infinite temperature bath with $\Gamma_\uparrow=\Gamma_\downarrow$.

The measurement was done when Qubit 2 was detuned from its symmetry point to a frequency $\nu_{ge} = 6.34$ GHz. Simulation was done for the same parameters as in Section \ref{sec:Numerical}, except for $\chi/2\pi = 0.003$. The system was solved for Eq. 3 of the main text, with no Rabi drive but including a variable decay rate $\Gamma_{\uparrow\downarrow}$.

\newpage

\section{Ramsey coherence times for different rifling angles}\label{sec:ramsey}

\begin{figure}[h!]
    \centering
    \includegraphics[width=0.75\linewidth]{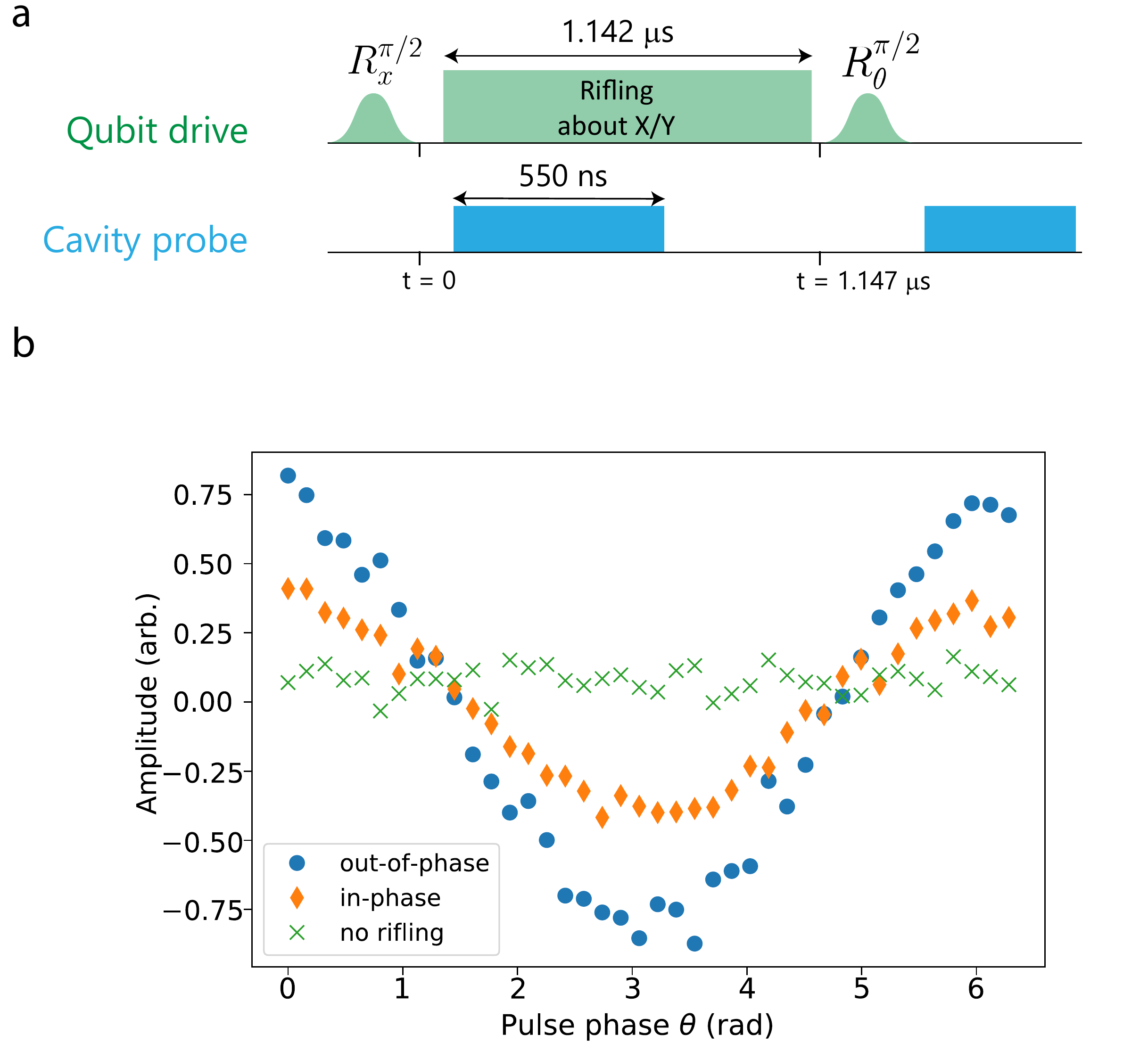}
    \caption{\textbf{Ramsey-type of experiment while rifling the qubit around different axes. a,} Protocol of the Ramsey-type experiment. \textbf{b, }Averaged result of the measurement described in (a) as a function of the phase $\theta$ of the second gaussian pulse. }
    \label{fig:ramsey}
\end{figure}

In this section we present measurements where we rifle Qubit 2 around two different axes. Following a $R_x^{\pi/2}$ pulse, the qubit can be effectively driven around two distinct axes: if the Rabi drive is in phase (around the x-axis), the qubit's state will rotate around the Bloch sphere alternating between states $\vert 0\rangle$ and $\vert 1\rangle$. If, however, the Rabi drive is out of phase with the first pulse (i.e. we rotate about the y-axis), the qubit's state will stay spin-locked around the  $(\vert 0\rangle +i\vert 1\rangle)/\sqrt{2}$ state.

The protocol is shown in Figure \ref{fig:ramsey}\textbf{a}. It resembles a Ramsey experiment, however instead of varying the time between two $R^{\pi/2}$ pulses as for standard Ramsey measurement, we vary the phase $\theta$ of the second $R^{\pi/2}$ pulse relative to the first. In between the $R^{\pi/2}$ pulses, the qubit is Rabi driven either around the x-axis (in-phase with the first $R_x^{\pi/2}$ pulse) or the y-axis (out of phase).

Results of the two different rifling axes and of the measurement without rifling are shown in Figure \ref{fig:ramsey}\textbf{b}. Without the rifling pulse, the cavity probe pulse suppresses the superposition of the qubit and no oscillations is seen as a function of the second pulse phase. However, superposition is maintained if the qubit is rifled, independent of the rifling axis. The amplitude variation in the measurement is greater when the qubit is spin-locked. This is probably because we did not take particular care to fine tune the rifling time between the $\pi/2$ pulses. This does not affect the experiment when the qubit is spin-locked, but when the rifling pulse is in-phase with the first pulse, the qubit's state may be driven further from the Bloch sphere's x-y plane, thus decreasing the visibility of the final measurement.

\newpage

\section{ Tomography data - Two-qubit density operator measurement}

\begin{figure}[h!]
    \centering
    \includegraphics[width=\linewidth]{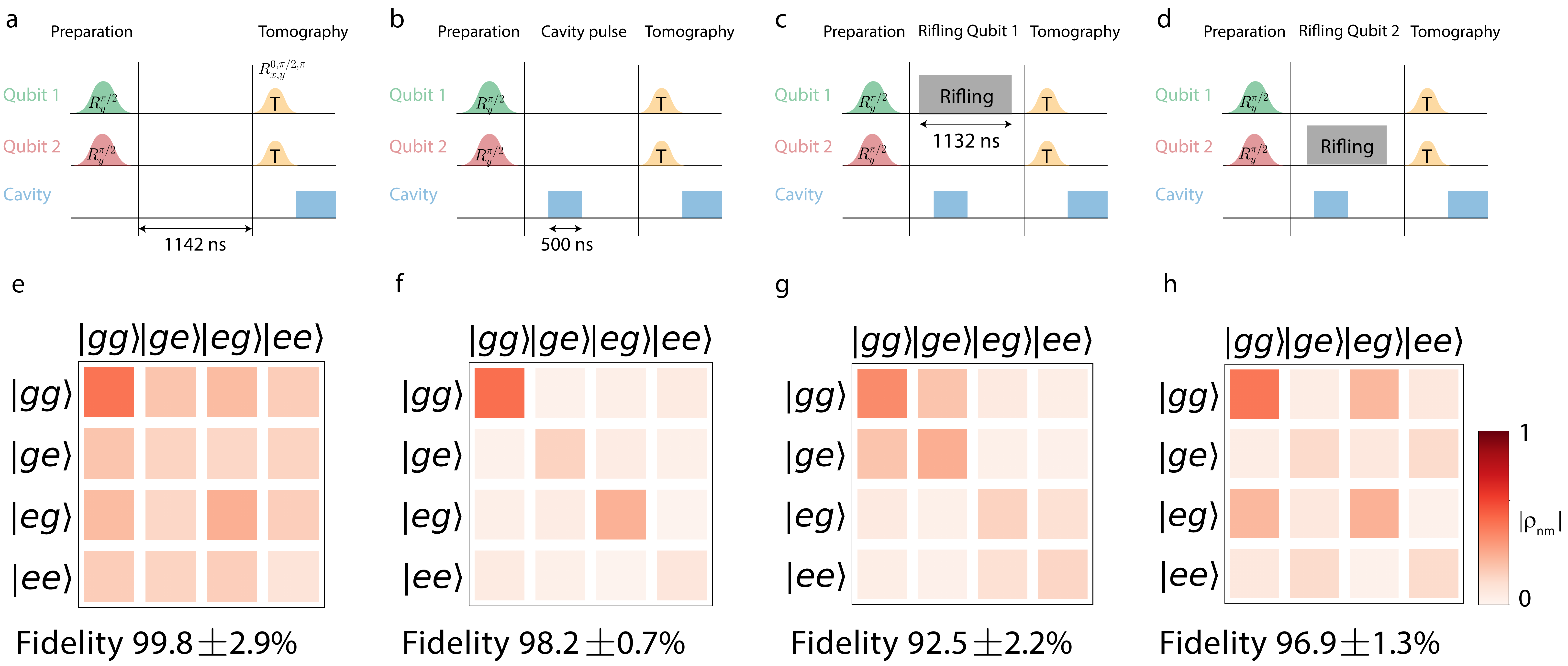}
    \caption{\textbf{Comparing tomography measurements of two qubits coupled to the same resonator with and without rifling. a - d, }Pulse protocol of each experiment. The last pulse on each qubit is either of the $R_{x,y}^{0,\pi/2,\pi}$ tomography pulses. \textbf{e-f, }Two qubit joint density matrices measured following each protocol depicted in (a)-(d). All matrices are extracted from integrating the measurement pulse after the tomography pulses. The fidelity values referenced are calculated by comparing to the ideal state assuming  an excited state decay for both qubits for 1142 ns. }
    \label{fig:two_qubit_tomo}
\end{figure}

In this section we present the results of tomography measurements of both Qubit 1 and Qubit 2 coupled to the same resonator. The setup was tuned to the same settings as those described for Figure 4 in the main text; as opposed to the single qubit density matrices in the main text, here we present joint density matrix measurements.

 The duration of the experiments, defined as the elapsed time between the first $R_y^{\pi/2}$ pulses used for preparation and the tomography pulses, is fixed to 1142 ns for all experiments.

In the first experiment (Fig. \ref{fig:two_qubit_tomo}a, e) we evaluate the joint state of both qubits in superposition, followed by a decay during the time of the experiment. In the second experiment (Fig. \ref{fig:two_qubit_tomo}b, f) we evaluate the effect of a 500 ns cavity pulse (with power corresponding to approximately 6.8 photons in the cavity, same as the second cavity pulse but shorter). As expected, the non-diagonal elements of the joint density matrix are negligible and we recover a fully classical state. In the third and fourth experiments  (Fig. \ref{fig:two_qubit_tomo}c-d, g-h) we rifle Qubit 1 and Qubit 2 respectively while leaving the other qubit idle. The rifling pulses are 1132 ns in duration and correspond in amplitude to a Rabi frequency of 30 MHz, and are applied out of phase with the first $R_y^{\pi/2}$ pulse such that the rifled qubit's state is spin locked (see section \ref{sec:ramsey} of this supplementary). For both cases we recover a joint density matrix corresponding to the rifled qubit in a superposition state while the other qubit's superposition has collapsed.

These measurements demonstrate that while a cavity pulse for read out destroys superposition, this can be preserved by rifling the appropriate qubit.

%\bibliography{references.bib}